\documentstyle[11pt]{article}

\makeatletter%
\def\nottoobig#1{{\hbox{$\left#1\vcenter to1.111\ht\strutbox{}\right.\n@space$}}}
\makeatother%

\makeatother%



\makeatletter \@beginparpenalty=10000 \makeatother
\def\desclabel#1{\bf #1\hfil}
\def\desc{\list{}{%
\labelwidth=\leftmargin
\advance \labelwidth by -\labelsep
\let \makelabel=\desclabel}}

\makeatletter %

\newcommand{\implies}{\:\Rightarrow\:}

\def\union{\,\bigcup\limits\,}

\newlength{\filength}
\newsavebox{\gcbox}
\sbox{\gcbox}{\framebox[\filength]{\rule{0ex}{2ex}}}

\newlength{\leftjustindent}
\newlength{\@leftjustindent}
\def\leftjust{\let\\\@leftjustcr\let\end\@endleftjust
  \addtolength{\@leftjustindent}{\leftjustindent}
  \vcenter\bgroup
  \halign\bgroup
    \hbox to\displaywidth{
      \rule{\@leftjustindent}{0ex}$\displaystyle##$\hfill
      }\crcr
}
\def\endleftjust{\crcr\egroup\egroup\endgroup}
\def\@endleftjust#1{\crcr\egroup\egroup\@checkend{#1}\endgroup}
\def\@leftjustcr{\crcr}

\newtheorem{theorem}{Theorem}[section]

\newtheorem{definition}[theorem]{Definition}

\newcommand{\qedblob}{\mbox{\rule[-1.5pt]{5pt}{10.5pt}}}
\def\literalqed{{\ \nolinebreak\hfill\mbox{\qedblob\quad}}}

\def\qed{\literalqed}

\newtheorem{lemma}[theorem]{Lemma}

\newcommand{\singlespacing}{\let\CS=
\@currsize\renewcommand{\baselinestretch}{1}\tiny\CS}
\newcommand{\singlespacingplus}{\let\CS=
\@currsize\renewcommand{\baselinestretch}{1.25}\tiny\CS}
\newcommand{\doublespacing}{\let\CS=
\@currsize\renewcommand{\baselinestretch}{1.75}\tiny\CS}
\newcommand{\draftspacing}{\let\CS=
\@currsize\renewcommand{\baselinestretch}{2.0}\tiny\CS}

\makeatother%

\singlespacingplus

\mathcode`\0="0030      %
\mathcode`\1="0031
\mathcode`\2="0032
\mathcode`\3="0033
\mathcode`\4="0034
\mathcode`\5="0035
\mathcode`\6="0036
\mathcode`\7="0037
\mathcode`\8="0038
\mathcode`\9="0039
\makeatletter
\clubpenalty=\@highpenalty
\widowpenalty=\@highpenalty
\makeatother

\makeatletter
\newcommand{\niceonespacing}{\let\CS=\@currsize\renewcommand{\baselinestretch}{1.1}\tiny\CS}\newcommand{\nicetwospacing}{\let\CS=\@currsize\renewcommand{\baselinestretch}{1.2}\tiny\CS}
\newcommand{\nicethreespacing}{\let\CS=\@currsize\renewcommand{\baselinestretch}{1.3}\tiny\CS}
\newcommand{\singlespacingplusplus}{\let\CS=\@currsize\renewcommand{\baselinestretch}{1.35}\tiny\CS}
\newcommand{\nicefourspacing}{\let\CS=\@currsize\renewcommand{\baselinestretch}{1.4}\tiny\CS}
\newcommand{\nicefivespacing}{\let\CS=\@currsize\renewcommand{\baselinestretch}{1.5}\tiny\CS}
\newcommand{\nicesixpacing}{\let\CS=\@currsize\renewcommand{\baselinestretch}{1.6}\tiny\CS}
\makeatother

\makeatletter
\def\@cite#1#2{[#1\if@tempswa , #2\fi]}
\makeatother

\makeatletter
\def\@citex[#1]#2{\if@filesw\immediate\write\@auxout{\string\citation{#2}}\fi
  \def\@citea{}\@cite{\@for\@citeb:=#2\do
    {\@citea\def\@citea{,\linebreak[0]}\@ifundefined
       {b@\@citeb}{{\bf ?}\@warning
       {Citation `\@citeb' on page \thepage \space undefined}}%
\hbox{\csname b@\@citeb\endcsname}}}{#1}}
\makeatother

\makeatletter
\def\ps@thesis{\def\@oddhead{\hfil\rm\thepage\hfil}\def\@oddfoot{}\def\@evenhead{\hfil\rm\thepage\hfil}\def\@evenfoot{}\def\chaptermark##1{}\def\sectionmark##1{}}
\makeatother

\newcommand{\p}{{\rm P}}

\newcommand{\littlepitalic}{{\it p}}

\newcommand{\np}{{\rm NP}}

\newcommand{\bh}{{\rm BH}}

\newcommand{\sigmak}{{{\rm \Sigma}_{k}^p}}

\newcommand{\sigmakplusone}{{\rm \Sigma}_{k+1}^p}
\newcommand{\sigmakmone}{{\rm \Sigma}_{k-1}^p}

\newcommand{\sigmai}{{\rm \Sigma}_i^p}
\newcommand{\sigmaipone}{{\rm \Sigma}_{i+1}^p}
\newcommand{\sigmaiptwo}{{\rm \Sigma}_{i+2}^p}

\newcommand{\pii}{{\rm \Pi}_i^p}

\newcommand{\wh}[1]{\widehat{#1}}
\newcommand{\whp}{{\wh{p}}}
\newcommand{\lsigk}{ {L_{\sigmak}}}

\newcommand{\lsigi}{ {L_{\sigmai}}}
\newcommand{\lpii}{ {L_{\pii}}}

\newcommand{\lsigipone}{ {L_{\sigmaipone}}}
\newcommand{\lsigiptwo}{ {L_{\sigmaiptwo}}}

\newcommand{\diffmsigk}{{\rm DIFF}_m(\sigmak)}
\newcommand{\diffssigi}{{\rm DIFF}_s(\sigmai)}

\newcommand{\diffspii}{{\rm DIFF}_s(\pii)}

\newcommand{\diff}{{\rm DIFF}}
\newcommand{\codiffmsigk}{{\rm co}{\diffmsigk}}

\newcommand{\ldiffmsigk}{{L_{\diffmsigk}}}
\newcommand{\ldiffssigi}{{L_{\diffssigi}}}
\newcommand{\ldiffspii}{{L_{\diffspii}}}

\newcommand{\deltatilde}{\tilde{\Delta}}
\newcommand{\bolddelta}{{\bf \Delta}}
\newcommand{\co}{{\rm co}}

\newcommand{\conp}{{\rm coNP}}

\newcommand{\ph}{{\rm PH}}

\def\pair#1{{{\langle\!\!~#1~\!\!\rangle}}}

\newcommand{\manyone}{\mbox{$\,\leq_{\rm m}^{{\littlepitalic}}$\,}}

\newcommand{\sigmastar}{\mbox{$\Sigma^\ast$}}

\newcommand{\calc}{{\cal C}}

\newcommand{\condition}{\,\nottoobig{|}\:}

\begin{document}
\sloppy

\title{Downward Collapse from 
a Weaker Hypothesis\protect\footnote{This research was supported in part 
by grants NSF-CCR-9322513 and
NSF-INT-9513368/DAAD-315-PRO-fo-ab, and was done in part during visits
to Le~Moyne College and to Friedrich-Schiller-Universit\"at Jena.}}

\author{Edith Hemaspaandra
\\
Department of Computer Science\\ Rochester Institute of 
Technology, \\ Rochester,
NY 14623\\
eh@cs.rit.edu
\and
Lane A. Hemaspaandra
\\Department of Computer Science\\ University of Rochester,\\ Rochester,
NY 14627\\
lane@cs.rochester.edu
\and
Harald Hempel\\
Institut f\"ur Informatik\\ Friedrich-Schiller-Universit\"at Jena,\\
07743 Jena, Germany\\
hempel@informatik.uni-jena.de}

\date{}

{\singlespacing 
\maketitle
}

{\singlespacing
\begin{abstract}
Hemaspaandra et al.~\cite{hem-hem-hem:tSPECIAL:translating-downwards}
proved that, for $m > 0$ and $0 < i < k - 1$: 
if $\sigmai \bolddelta \diffmsigk$ is closed under complementation,
then $\diffmsigk = \codiffmsigk$.
This sharply asymmetric result
fails to apply to the 
case in which 
the hypothesis is weakened by allowing the $\Sigma_i^p$ to 
be replaced by any class in its difference hierarchy.
We so extend the result by proving that,
for $s,m > 0$ and $0 < i < k - 1$:
if $\diffssigi \bolddelta \diffmsigk$ is closed under complementation,
then $\diffmsigk = \codiffmsigk$.
\end{abstract}
} %

\section{Introduction and Preliminaries}

In complexity theory, countless cases are known in which it can be
proven that the collapse of seemingly small classes implies the
collapse of classes that before-the-fact seemed potentially larger,
e.g., $\np = \conp \implies \np = \ph$.  Such theorems are known as
``upward collapse'' results, as they translate equalities 
(seemingly) upwards.
Upward collapse results for the polynomial hierarchy date back to the
1970s---in particular, to the classic 
papers that introduced the polynomial hierarchy~\cite{mey-sto:b:reg-exp-needs-exp-space,sto:j:poly}.

However, nontrivial {\em downward\/} translation-of-equality 
results---so-called 
``downward collapse'' results---within the polynomial
hierarchy's bounded query levels are a very recent attainment.  The
first was obtained by Hemaspaandra et
al.~\cite{hem-hem-hem:j:downward-translation}, who showed, with $k>2$,
that if one and two queries to $\sigmak$ yield the same computational
power, then the polynomial hierarchy collapses to $\sigmak$.  That
result was extended to other cases by Buhrman and
Fortnow~\cite{buh-for:c:two-queries} and Hemaspaandra et
al.~\cite{hem-hem-hem:tSPECIAL:translating-downwards}.  Interesting related
work has been done by
Wagner~\cite{wag:j:parallel-difference,rei-wag:ctoappear:boolean-lowness}
and others.  Hemaspaandra et
al.~\cite{hem-hem-hem:tWsurveyInTitle:easy-hard-survey} have written a
survey of the active history of this research area, all of which can
be viewed as sharply extending the power of Kadin's important
``easy-hard technique''~\cite{kad:joutdatedbychangkadin:bh}.  

The underlying motivation behind the study of downward collapse
results is to---via combining them with upward collapse results---show
that collapses that seemed to be different issues in fact are 
the same issue in disguise.
This paper does exactly that.
In particular, 
this paper establishes a new downward 
separation result extending the following theorem proven by 
Hemaspaandra et al.~\cite{hem-hem-hem:tSPECIAL:translating-downwards}:
For $m > 0$ and $0 < i < k - 1$:
if $\sigmai \bolddelta \diffmsigk$ is closed under complementation,
then $\diffmsigk = \codiffmsigk$.  However, since the converse direction
is trivial, our result yields the following link:
\begin{quote}For each $m > 0$ and $0 < i < k - 1$:
$\sigmai \bolddelta \diffmsigk$ is closed under complementation
if and only if
$\diffmsigk = \codiffmsigk$.
\end{quote}

In a moment we will define the key notations that may be 
unfamiliar to the reader, but in case of notational questions, we 
note that 
all notations are as in 
Hemaspaandra et al.~\cite{hem-hem-hem:tSPECIAL:translating-downwards}, 
and also the proof here is 
based on extending
Hemaspaandra et al.'s~proof combined with showing that a
lemma of Beigel,
Chang, and Ogihara~\cite{bei-cha-ogi:j:difference-hierarchies}
applies to prefixes of $\sigmastar$.
So as to make maximally clear to the reader familiar with
both the points of difference,
we exactly follow when possible the wording and structure 
of~\cite{hem-hem-hem:tSPECIAL:translating-downwards},
except in those places where this proof must diverge from
that proof in order to obtain its clearly stronger result.

We 
now state some standard definitions, and a useful lemma
from~\cite{hem-hem-hem:tSPECIAL:translating-downwards}. The ``$\Delta$''
classes mentioned in this definition are important throughout the
research on easy-hard-technique-based downward collapses.  In
particular, Selivanov~(see
\cite{sel:j:fine-hierarchies,sel:c:refined-ph}) shows that if such
classes are closed under complementation, the polynomial hierarchy
collapses.  This might already seem to yield our result, but it
does not.  Selivanov (under the complementation
hypothesis) collapses the polynomial hierarchy to a level 
containing $\sigmakplusone$, and thus shows merely
an upward translation of equality. 
In contrast, we collapse the difference hierarchy over $\sigmak$ to a level 
that is contained in the classes of the complementation hypothesis---thus
obtaining a new
downward translation of equality. 
Also, we note that our main theorem 
implies a collapse of the polynomial 
hierarchy to a class a full level lower in the difference hierarchy over 
$\sigmakplusone$ than could be concluded without our downward collapse 
result~(namely to 
$\diff_m(\sigmak)\bolddelta
\diff_{m-1}
(\sigmakplusone)$, in light of the strongest known
``BH/PH-collapse connection,'' 
see
\cite{hem-hem-hem:tWsurveyInTitle:easy-hard-survey,rei-wag:ctoappear:boolean-lowness}).

\begin{definition}
\begin{enumerate}
\item
For any classes ${\cal C}$ and ${\cal D}$, 
\[{\cal C} \bolddelta {\cal D} = \{L \condition (\exists C \in {\cal C})
(\exists D \in {\cal D}) [ L = C \Delta D]\},\]
where $C \Delta D = (C - D) \cup (D - C)$.
\item
For any sets ${C}$ and ${D}$,
\[C \deltatilde D =
\{\pair{x, y} \condition x \in C \Leftrightarrow y \not
\in D\}.\]
\item
(\cite{cai-gun-har-hem-sew-wag-wec:j:bh1,cai-gun-har-hem-sew-wag-wec:j:bh2},
see also~\cite{hau:b:sets,koe-sch-wag:j:diff}) \quad Let $\cal C$ be any
complexity class.  We now define the levels of  the boolean hierarchy.
\begin{enumerate}
\item $ {\rm DIFF}_1(\calc) = \calc$.
\item For any $k \geq 1$, ${\rm DIFF}_{k+1}(\calc) =
\{ L \condition  (\exists L_1 \in \calc)(\exists L_2 \in 
{\rm DIFF}_k(\calc))[ L = L_1 - L_2]\}$.
\item For any $k \geq 1$, $ {\rm coDIFF}_k(\calc) = 
\{ L \condition \overline{L} \in 
{\rm DIFF}_k(\calc)\}$.
\item $\bh(\calc)$, 
the boolean hierarchy over $\calc$, is 
$\union_{k \geq 1\,} {\rm DIFF}_k$.
\end{enumerate}
\end{enumerate}
\end{definition}

\begin{lemma}[\cite{hem-hem-hem:tSPECIAL:translating-downwards}]
\label{l:tildecomplete}
$C$ is $\manyone$-complete for ${\cal C}$ and
$D$ is $\manyone$-complete for ${\cal D}$, then
$C \deltatilde D$ is $\manyone$-hard for ${\cal C} \bolddelta {\cal D}$.
\end{lemma}

\begin{definition}\label{d:languages}
Let $\lsigi$, $\lsigipone$, and $\lsigiptwo$ be  $\manyone$-complete  languages
for 
$\sigmai$, $\sigmaipone$, and $\sigmaiptwo$, respectively that
satisfy\footnote{By 
the Stockmeyer-Wrathall~\cite{sto:j:poly,wra:j:complete} 
quantifier characterization
of the polynomial hierarchy's levels, such sets do exist.}
$$\lsigipone = \{x\condition (\exists y \in \Sigma^{|{x}|})[\pair{x,y} 
\notin \lsigi]\},$$
and 
$$\lsigiptwo = \{x\condition (\exists y\in \Sigma^{|{x}|})
[\pair{x,y} \notin \lsigipone]\}.$$

Let $\lsigk$ be a  $\manyone$-complete  language for $\sigmak$ and
let $\ldiffmsigk$ be $\manyone$-complete 
for $\diffmsigk$.
Let $\lpii=\overline{\lsigi}$ and define 
$L_{\diff_1(\pii)}=\lpii$ and for $j\geq 2$,  
$L_{\diff_j(\pii)}=\{\pair{x,y}\condition x \in \lpii \wedge y \notin 
L_{\diff_{j-1}(\pii)}\}$.
Note that $L_{\diff_j(\pii)}$ is many-one complete for $\diff_j(\pii)$ 
for all $j \geq 1$.
Note that $\diff_j(\pii)=\diff_j(\sigmai)$ if $j$ is even and 
$\diff_j(\pii)=\co\diff_j(\sigmai)$ if $j$ is odd.
Let $\ldiffssigi=\ldiffspii$ if $s$ is even and 
$\ldiffssigi=\overline{\ldiffspii}$ if $s$ is odd.
Then $\ldiffssigi$ is  $\manyone$-complete  for  
$\diffssigi$.
\end{definition}

Finally, we mention in passing that the study of downward collapse
results is closely related to the study of the power of query 
order---whether the order in which databases are accessed matters---an 
area recently introduced by Hemaspaandra, Hempel, and
Wechsung~\cite{hem-hem-wec:jtoappear:query-order-bh}.  In 
particular, downward collapse techniques have been used to understand
the power of query order within the polynomial
hierarchy (see~\cite{hem-hem-hem:j:query-order-ph}, the 
survey~\cite{hem-hem-hem:j:query-order-survey}, and the 
references therein, especially~\cite{wag:j:parallel-difference}).

\section{Main Result}

We now turn to the main result.

\begin{theorem}\label{t:newrest}
Let $s,m > 0$ and $0 < i < k - 1$.
If $\diffssigi \bolddelta \diffmsigk$ is closed under complementation,
then $\diffmsigk = \codiffmsigk.$
\end{theorem}

\noindent
{\bf Proof of Theorem~\ref{t:newrest}} \quad 
Since for $s=1$ this is exactly the 
main claim of
\cite[Section~3]{hem-hem-hem:tSPECIAL:translating-downwards}, 
we
henceforward assume 
$s\geq 2$.
Since $\ldiffssigi \deltatilde \ldiffmsigk$ is $\manyone$-hard for
 $\diffssigi \bolddelta \diffmsigk$ by Lemma~\ref{l:tildecomplete} 
(in fact, it is not hard to see that it even is $\manyone$-complete
for that class) and by
assumption
$\diffssigi \bolddelta \diffmsigk$ is closed under complementation, 
there exists a
polynomial-time many-one reduction $h$ from
$\ldiffssigi \deltatilde \ldiffmsigk$ to its complement.
That is, for all $x_1, x_2 \in \sigmastar$ it holds that:
if $h(\pair{x_1, x_2}) = \pair{y_1, y_2}$, 
then:
$\pair{x_1, x_2} \in \ldiffssigi \deltatilde \ldiffmsigk 
\Leftrightarrow \pair{y_1, y_2} \not \in
\ldiffssigi \deltatilde \allowbreak \ldiffmsigk$. 
Equivalently,
for all $x_1, x_2 \in \sigmastar$:
\begin{quotation}
\noindent
{\bf Fact~1:} \\
if $h(\pair{x_1, x_2}) = \pair{y_1, y_2}$, then:
\begin{eqnarray*}
\lefteqn{(x_1 \in \ldiffssigi \Leftrightarrow x_2 \notin \ldiffmsigk)
\mbox{ if and only if }}\\
&&(y_1 \in \ldiffssigi \Leftrightarrow y_2 \in \ldiffmsigk).
\end{eqnarray*}
\end{quotation}
We can use $h$ to recognize some of $\overline{\ldiffmsigk}$ by a $\diffmsigk$
algorithm. 
In particular, 
we say that a string $x$ is {\em easy for length $n$\/} 
if there exists a string $x_1$ such that
$|x_1| \leq n$ and $(x_1 \in \ldiffssigi
\Leftrightarrow y_1 \in \ldiffssigi)$ where
$h(\pair{x_1, x}) = \pair{y_1, y_2}$.

Let $p$ be a fixed polynomial, which will be exactly 
specified later in the proof.  We have the following algorithm to test whether
$x \in \overline{\ldiffmsigk}$ in the 
case that (our input) $x$ is an easy string
for $p(|x|)$.
On input $x$, guess $x_1$ with $|x_1| \leq p(|x|)$, let
$h(\pair{x_1, x}) = \pair{y_1, y_2}$,
and accept if and only if
$(x_1 \in \ldiffssigi \Leftrightarrow y_1 \in \ldiffssigi)$  and
$y_2 \in \ldiffmsigk$.  
This algorithm is not necessarily a $\diffmsigk$ algorithm, 
but it does inspire the following 
$\diffmsigk$ algorithm  to test whether $x \in \overline{\ldiffmsigk}$ in the
case that $x$ is an easy string for $p(|x|)$.

Let $L_1, L_2, \cdots, L_m$ be languages in $\Sigma^p_k$ such that
$\ldiffmsigk = L_1 - (L_2 - (L_3 - \cdots  (L_{m-1} - L_m) \cdots))$
and $L_1 \supseteq L_{2} \supseteq \cdots \supseteq L_{m-1}
\supseteq L_m$
(this can be done, as it is simply
the ``telescoping'' normal form of the levels of the 
boolean hierarchy over $\sigmak$, see
\protect\cite{cai-gun-har-hem-sew-wag-wec:j:bh1,hau:b:sets,wec:c:bh:ormaybe:wech:only:is:right}).
For $1 \leq r \leq m$, define $L_r'$ as the language accepted by the
following 
$\sigmak$ machine: On input $x$, guess $x_1$ with $|x_1| \leq p(|x|)$,
let $h(\pair{x_1, x}) = \pair{y_1, y_2}$,
and accept if and only if
$(x_1 \in \ldiffssigi \Leftrightarrow y_1 \not \in \ldiffssigi)$ and $y_2 \in
L_r$.

Note that $L'_r \in \sigmak$
for each $r$, and that $L'_1 \supseteq L'_{2} \supseteq \cdots
\supseteq L'_{m-1}  \supseteq L'_m$. 
We will show that if $x$ is an easy string for
length $p(|x|)$, then $x \in \overline{\ldiffmsigk}$ if and only if
$x \in L'_1 - (L'_2 - (L'_3 - \cdots  (L'_{m-1} - L'_m) \cdots))$.

So suppose that $x$ is an easy string for $p(|x|)$.
Define  $r'$
to be the unique
integer
such that
(a)~$0 \leq r' \leq m$,
(b)~$x \in L'_{s}$ for $1 \leq s \leq r'$,
and (c)~$x \not \in L'_{s}$ for $s > r'$.  It is immediate that  
$x \in L'_1 - (L'_2 - (L'_3 - \cdots  (L'_{m-1} - L'_m)
\cdots))$ if and only if $r'$ is odd.

Let $w$ be some string such that:
\begin{itemize}
\item $(\exists x_1 \in 
(\sigmastar)^{\leq p(|x|)}) (\exists y_1) [h(\langle x_1,
x \rangle ) = 
\langle y_1, w \rangle 
\wedge  (x_1 \in \ldiffssigi \Leftrightarrow y_1 \not \in
\ldiffssigi)]$, and
\item $w \in L_{r'}$ if $r' > 0$.
\end{itemize}
Note that such a $w$ exists, since $x$ is easy for $p(|x|)$. 
By the definition of $r'$ (namely, since $x \not\in L'_s$
for $s>r'$), 
$w \not \in L_s$ for all
$s > r'$. It follows that $w \in \ldiffmsigk$ if and only if $r'$ is odd.

It is clear, keeping in mind the definition of $h$, that $x 
\in \overline{\ldiffmsigk}$ iff $w \in \ldiffmsigk$, 
$w \in \ldiffmsigk$ iff 
$r'$ is odd, and
$r'$ is odd iff $x \in L'_1 - (L'_2 - (L'_3 - \cdots 
(L'_{m-1} - L'_m) \cdots))$. 
This completes the case where $x$ is easy,
as 
$L'_1 - (L'_2 - (L'_3 - \cdots 
(L'_{m-1} - L'_m) \cdots))$ in effect specifies 
a $\diffmsigk$ algorithm.

We say that $x$ is {\em hard for length $n$\/} if 
$|x| \leq n$ and $x$ is not easy for length $n$, i.e., if
$|x| \leq n$ and, for all $x_1$ with $|x_1| \leq n$, $(x_1 \in \ldiffssigi
\Leftrightarrow y_1 \notin \ldiffssigi)$, where
$h(\pair{x_1, x}) = \pair{y_1, y_2}$.

If $x$ is a hard string for length $n$, then $x$ induces a
many-one reduction from 
${\left(\ldiffssigi\right)}^{\leq n}$ to $\overline{\ldiffssigi}$,
namely, $f_x(x_1) = y_1$, where $h(\pair{x_1, x}) = \pair{y_1, y_2}$.
Note that there is a 
particular polynomial-time function that 
simultaneously  implements all
the $f_x$, namely the function $\widehat{a}(x,x_1) = y_1$, where $h(\pair{x_1,
x}) = \pair{y_1, y_2}$ provides such. Henceforward, we will speak of $f_x$, and
similar notions, and will take as tacit the fact that they,
similarly, are uniformly
implementable.

It is known
that a collapse of the boolean hierarchy over 
$\sigmai$ implies a collapse of the polynomial 
hierarchy~\cite{cha-kad:j:closer,bei-cha-ogi:j:difference-hierarchies}. 
A long series of papers studied the question to what level the polynomial 
hierarchy collapses in that case. 
The best known 
results (\cite{cha-kad:j:closer,bei-cha-ogi:j:difference-hierarchies,hem-hem-hem:j:downward-translation,rei-wag:ctoappear:boolean-lowness,hem-hem-hem:tWsurveyInTitle:easy-hard-survey}, 
see especially the strongest such connection, which is that
obtained independently in 
\cite{rei-wag:ctoappear:boolean-lowness} and
\cite{hem-hem-hem:tWsurveyInTitle:easy-hard-survey})
conclude 
a collapse of the polynomial hierarchy to a level within the boolean 
hierarchy over $\Sigma^p_{i+1}$. 
Though a hard string for length $n$ only induces a many-one reduction 
between initial segments of
$\ldiffssigi$ and $\overline{\ldiffssigi}$, we would nevertheless 
like to derive at least 
a $\p^{\sigmakmone}$ algorithm for some of $\lsigiptwo$. 
The following lemma does exactly that. 

\begin{lemma}\label{l:hhh}
Let $s > 1$, $m >0$, and $0 <i < k - 1$,
and suppose that
$\diffssigi\bolddelta\diffmsigk=\co(\diffssigi\bolddelta\diffmsigk)$.
There exist a set $D \in \p^{\sigmaipone}$ 
and a polynomial $r$ such that for all $n$, 
(a) $r(n+1) > r(n) > 0$ and (b) for all $x \in \sigmastar$, 
if $x$ is a hard string for length $r(n)$ 
then for all $y \in (\sigmastar)^{\leq n}$,
$$y\in \lsigiptwo \iff \pair{x,1^n,y} \in D.$$
\end{lemma}

We defer the proof of Lemma~\ref{l:hhh} and first finish the proof of 
our theorem.

If $x$ is a hard string for length $p(|x|)$ 
we will use the result of  
Lemma~\ref{l:hhh} 
to obtain a $\p^{\sigmakmone}$ algorithm 
for some of $\ldiffmsigk$, and hence 
(since $\p^{\sigmakmone}$ is closed under complementation)  
certainly a $\p^{\sigmakmone}$ algorithm for
some of $\overline{\ldiffmsigk}$.

To be more precise, suppose that $x$ is a hard string for length $r(n)$. 
According to the above Lemma~\ref{l:hhh}, $x$ induces a 
$\p^{\sigmaipone}$ algorithm for all strings in 
${\left(\lsigiptwo\right)}^{\leq n}$ that runs in time polynomial in $n$.
What we would like to conclude is a $\p^{\sigmakmone}$ 
algorithm for ${\left(\ldiffmsigk\right)}^{= |x|}$.
Recall that 
$\ldiffmsigk = L_1 - (L_2 - (L_3 - \cdots  (L_{m-1} - L_m) \cdots))$, 
where $L_j \in \sigmak$ for all $j$. 
Since $\lsigk$ is complete for $\sigmak$, there exist functions 
$f_1,\cdots,f_m$ which many-one reduce $L_1,\cdots,L_m$ to $\lsigk$, 
respectively.
Let the output sizes of all the $f_j$'s be bounded by the polynomial $p'$,  
which without loss of generality satisfies $(\forall 
\widehat{m} \geq 0)[p'(\widehat{m}+1) > p'(\widehat{m}) > 0]$.  
Hence 
an $x$-induced $\p^{\Sigma^p_{k-1}}$ algorithm for strings in 
${\left(\lsigk\right)}^{\leq p'(|x|)}$ suffices to give us a 
$\p^{\Sigma^p_{k-1}}$ algorithm for strings in 
${\left(\ldiffmsigk\right)}^{= |x|}$. 
But Lemma~\ref{l:hhh} gives us exactly this, if $k=i+2$ and if $x$ is hard 
for length $r(p'(|x|))$.
For the case $k>i+2$, let $M$ be a $\Sigma^p_{k-(i+2)}$ machine recognizing 
$\lsigk$ 
with oracle queries to $\lsigiptwo$ and running in time $q'$ for some 
polynomial $q'$ satisfying  $(\forall 
\widehat{m} \geq 0)[q'(\widehat{m}+1) > q'(\widehat{m}) > 0]$.
We can certainly replace the $\lsigiptwo$ queries by queries to a 
$\p^{\sigmaipone}$ oracle and 
thus obtain a $\Sigma^p_{k-1}$ algorithm 
(running in time polynomial in $|x|$) 
for $\left(\lsigk\right)^{\leq p'(|x|)}$, 
if we ensure that Lemma~\ref{l:hhh} 
gives us an $x$-induced $\p^{\sigmaipone}$ algorithm for all strings in 
${\left(\lsigiptwo\right)}^{\leq q'(p'(|x|))}$. 
Thus, if $k>i+2$ we need $x$ to be hard for length $r(q'(p'(|x|)))$. 

So let $p$ be an easily computable polynomial satisfying  $(\forall 
\widehat{m} \geq 0)[p(\widehat{m}+1) > p(\widehat{m}) > 0]$ and for all $n$,
$p(n) \geq r(q'(p'(n)))$ ($p(n) \geq r(p'(n))$) if $k>i+2$ ($k=i+2$). 
As promised, we now have specified $p$.

However, now we have an outright $\diffmsigk$ algorithm 
for $\overline{\ldiffmsigk}$: 
For $1 \leq r \leq m$ define a
$\np^{\Sigma_{k-1}^p}$  machine  $N_r$ as follows:
On input $x$, the NP base machine of $N_r$ 
executes the following algorithm:
\begin{enumerate}
\item Using its 
$\Sigma_{k-1}^p$ oracle, it deterministically 
determines whether the input $x$ is an easy string for 
length $p(|x|)$.  This can be done, as checking whether 
the input is an easy string for length $p(|x|)$ can be done 
by one query to $\Sigma_{i+1}^p$, and $i+1 \leq k-1$ by 
our $i < k-1$ hypothesis.
\item If the previous step determined that the input is not 
an easy string, then the input must be a hard string
for length $p(|x|)$.  
If $r=1$ then simulate the $\p^{\Sigma^p_{k-1}}$ algorithm for 
$\overline{\ldiffmsigk}$
induced by this hard string
(i.e., the input $x$ itself) on input $x$ (via our NP
machine itself simulating the base $\p$ machine of the 
$\p^{\Sigma^p_{k-1}}$ algorithm and using the NP machine's oracle to 
simulate the oracle queries made by the base P machine 
of the 
$\p^{\Sigma^p_{k-1}}$ algorithm being simulated).
If $r>1$ then reject.
\item If the first step determined 
that the input $x$ is easy for length $p(|x|)$, then our NP
machine
simulates (using itself and its oracle) 
the $\sigmak$ algorithm for  $L'_r$ on input $x$.
\end{enumerate}
Note that the 
$\Sigma_{k-1}^p$ oracle in the above algorithm is being used
for a number of different sets.  However, as 
$\Sigma_{k-1}^p$ is closed under disjoint 
union, this presents no 
problem as we can use the disjoint union of the sets, 
while modifying the queries so they address the 
appropriate part of the disjoint union.

It follows that, for all $x$, $x  
\in \overline{\ldiffmsigk}$ if and only if
$x \in L(N_1) - (L(N_2) - (L(N_3) - \cdots 
(L(N_{m-1}) - L(N_m)) \cdots))$.
Since $\overline{\ldiffmsigk}$ is complete for $\codiffmsigk$,
it follows that  $\diffmsigk = \codiffmsigk$.~\qed

We now give the proof of Lemma~\ref{l:hhh}. 
The upcoming proof should be seen in the context with the proof of 
Theorem~\ref{t:newrest} as some notations we are going to use are 
defined there.

\smallskip

\noindent
{\bf Proof of Lemma~\ref{l:hhh}} \quad 
Our proof 
follows and generalizes
a proof from~\cite{bei-cha-ogi:j:difference-hierarchies}. 
Let $\langle \cdots \rangle$ be a pairing function that maps sequences of 
length at most $2s+2$ of strings over $\sigmastar$ to $\sigmastar$ having 
the standard properties such as polynomial-time computability and 
invertibility, etc.
Let $t$ be a polynomial such that 
$|\pair{x_1,x_2,\dots,x_j}| \leq t(\max\{|x_1|,|x_2|,\dots ,|x_j|\})$ 
for all $1\leq j\leq 2s + 2$ and all $x_1,x_2,\dots,x_j \in \sigmastar$. 
Without loss of generality let $t$ be such that $t(n+1)>t(n)>0$ for all $n$.
Define $t^{(0)}(n)=n$ and 
$t^{(j)}(n)=\underbrace{t(t(\cdots t}_{j~times}(n) \cdots ))$ for all $n$ 
and all $j\geq 1$.

Define $r$ to be a polynomial such that 
$r(n+1) > r(n) > 0$ and $r(n)\geq t^{(s-1)}(n)$ for all $n$.
Let $n$ be an integer.
Suppose that $x$ is a hard string for length $r(n)$ 
as defined in the proof of Theorem~\ref{t:newrest}.
Then, for all $y$ such that $|y|\leq r(n)$,
$$y \in \ldiffssigi \iff f_x(y) \not\in \ldiffssigi,$$
or equivalently
$$y \in \ldiffspii \iff f_x(y) \not\in \ldiffspii.$$

Recall that $f_x(y)$ can be computed in time polynomial in $\max\{|x|,|y|\}$.
Let $y=\pair{y_1,y_2}$ and let $f_x(y)=\pair{z_1,z_2}$.
Then, for all $y_1,y_2 \in \sigmastar$ such that $|y_1|\leq n$ and 
$|y_2| \leq t^{(s-2)}(n)$, 
$$(*)\qquad y_1 \in \lpii \wedge y_2 \not\in L_{\diff_{s-1}(\pii)} \iff
z_1 \not\in \lpii \vee z_2 \in  L_{\diff_{s-1}(\pii)}.$$

We say that $y_1$ is $s$-easy for length $n$ if and only if $|y_1|\leq n$ and 
$(\exists y_2~|y_2|\leq t^{(s-2)}(n))[z_1 \not\in \lpii]$.
$y_1$ is said to be $s$-hard for length $n$ if and only if $|y_1|\leq n$, 
$y _1 \in \lpii$, and 
$(\forall y_2~|y_2|\leq t^{(s-2)}(n))[z_1 \in \lpii]$.
Observe that the above notions are defined with respect to our hard string 
$x$, since $z_1$ depends on $x$, $y_1$, and $y_2$.
Furthermore, according to (*), if $y_1$ is $s$-easy for length $n$ 
then $y_1 \in \lpii$.

Suppose there exists an $s$-hard string $\omega_s$ for length $n$. 
Let $f_{(x,\omega_s)}$ be the function defined by 
$f_x(\pair{\omega_s,y})=\pair{z_1, f_{(x,\omega_s)}(y)}$.
$f_{(x,\omega_s)}(y)$ can be computed in time polynomial in 
$\max\{|x|,|\omega_s|,|y|\}$.
In analogy to the above we define $(s-1)$-easy and $(s-1)$-hard
strings. If an $(s-1)$-hard string exists we can repeat the process and define 
$(s-2)$-easy and $(s-2)$-hard strings and so on.
Note that the definition of $j$-easy and $j$-hard strings can only be made 
with respect to our hard string $x$, some fixed $s$-hard string $\omega_s$, 
some fixed $(s-1)$-hard string $\omega_{s-1}$, \dots, 
some fixed $(j+1)$-hard string $\omega_{j+1}$.
If we have found a sequence of strings 
$(\omega_s,\omega_{s-1},\dots,\omega_2)$ such that every $\omega_j$ is 
$j$-hard with respect to $(x,\omega_s,\omega_{s-1},\dots,\omega_{j+1})$ 
then we have for all $y$, $|y|\leq n$,
$$y \in \lpii \iff f_{(x,\omega_s,\omega_{s-1},\dots,\omega_2)}(y) 
\notin \lpii.$$
We say that a string $y$ is 1-easy for length $n$ if and only if 
$|y| \leq n$ and 
$f_{(x,\omega_s,\omega_{s-1},\dots,\omega_2)}(y) \notin \lpii$. 
We define that no string is 1-hard for length $n$.

$(x)$ is called a hard sequence for length $n$.
A sequence $(x,\omega_s,\omega_{s-1},\dots,\omega_j)$ of strings is called a 
hard sequence for length $n$ if and only if 
$\omega_s$ is $s$-hard with respect to $x$ and 
for all $\ell$, $j \leq \ell\leq s-1$, 
$\omega_{\ell}$ is $\ell$-hard with respect to 
$(x,\omega_s,\omega_{s-1},\dots,\omega_{\ell +1})$.
Note that given a hard sequence $(x,\omega_s,\omega_{s-1},\dots,\omega_j)$, 
the strings in $(\lpii)^{\leq n}$ divide into $(j-1)$-easy and 
$(j-1)$-hard strings 
(with respect to $(x,\omega_s,\omega_{s-1},\dots,\omega_j)$) 
for length $n$.

$(x)$ is called a maximal hard sequence if and only if there exists no 
$s$-hard string for length $n$.
A hard sequence $(x,\omega_s,\omega_{s-1},\dots,\omega_j)$ is called a 
maximal hard sequence for length $n$ if and only if 
there exists no $(j-1)$-hard string 
for length $n$ with respect to $(x,\omega_s,\omega_{s-1},\dots,\omega_j)$.
If we in the following denote a maximal hard sequence by 
$(x,\omega_s,\omega_{s-1},\dots,\omega_j)$ we explicitly include the 
case that the maximal hard sequence might be $(x)$ or 
$(x,\omega_s)$. 

{\bf Claim 1:} {\em There exists a set $A\in \sigmai$ such that if 
$(x,\omega_s,\omega_{s-1},\dots,\omega_j)$ is a maximal hard sequence 
for length $n$ then for all $y$ and $n$ satisfying $|y|\leq n$ it 
holds that:
$$y \in \lpii \iff 
\pair{x,1^n,\omega_s,\omega_{s-1},\dots,\omega_j,y} \in A.$$}

{\em Proof of Claim 1:} 
Let $(x,\omega_s,\omega_{s-1},\dots,\omega_j)$ be a maximal hard sequence 
for length $n$. Note that $j \geq 2$ and that
the
strings in $(\lpii)^{\leq n}$ are exactly the strings of length at most $n$
that are $(j-1)$-easy with  respect to
$(x,\omega_s,\omega_{s-1},\dots,\omega_j)$.  It is immediate
from the definition that testing  whether a string $y$ is $(j-1)$-easy for
length $n$ with  respect to $(x,\omega_s,\omega_{s-1},\dots,\omega_j)$ can be
done by a 
$\sigmai$ algorithm running in time polynomial in $n$: 
If $j \geq 3$, check $|y|\leq n$, guess $y_2$, 
$|y_2|\leq t^{(j-3)}(n)$, compute 
$f_{(x,\omega_s,\omega_{s-1},\dots,\omega_j)}(\pair{y,y_2})=\pair{z_1,z_2}$,
 and accept if and only if $z_1 \not\in \lpii$; If $j = 2$, check $|y|\leq n$,
and accept if and only if $f_{(x,\omega_s,\omega_{s-1},\dots,\omega_2)}(y)
\notin \lpii$.

{\bf Claim 2:} {\em There exist a set $B\in \sigmai$ and a 
 polynomial  $\whp$ such that  $(\forall 
n \geq 0)[\whp(n+1) > \whp(n) > 0]$ and if 
$(x,\omega_s,\omega_{s-1},\dots,\omega_j)$ is a maximal hard sequence 
for length $\whp(n)$ then for all $y$ and $n$
satisfying $|y|\leq n$ it holds that:
$$y \in \lsigipone \iff 
\pair{x,1^n,\omega_s,\omega_{s-1},\dots,\omega_j,y} \in B.$$}

{\em Proof of Claim 2:} 
Let $A \in \sigmai$ as in Claim 1.
Let $y$ be a string such that $|y|\leq n$.
According to the definition of $\lsigipone$,  
$$y \in \lsigipone \iff 
(\exists z \in \Sigma^{|y|})[\pair{y,z}\notin \lsigi].$$
Recall that $\lpii=\overline{\lsigi}$. 
Define $\whp$ to be a polynomial such that 
$\whp(n+1) >\whp(n)>0$ and $\whp(n)\geq t(n)$ for
all $n$. In light of Claim 1 we obtain that if 
$(x,\omega_s,\omega_{s-1},\dots,\omega_j)$ is a maximal hard sequence 
for length $\whp(n)$ then 
$$y \in \lsigipone \iff 
(\exists z \in \Sigma^{|y|})
[\pair{x,1^{\whp(n)},\omega_s,\omega_{s-1},\dots,\omega_j,\pair{y,z}}\in A].$$
We define
$B$ to be the set  
$B=\{\pair{x,1^n,\omega_s,\omega_{s-1},\dots,\omega_j,y}\condition 
(\exists z \in \Sigma^{|y|})
[\langle x,1^{\whp(n)}, \allowbreak
\allowbreak \omega_s,\omega_{s-1},\dots,\omega_j,\pair{y,z}\rangle\in A]\}.$
Clearly $B \in \sigmai$.
This proves the claim.

{\bf Claim 3:} {\em There exist a set $C\in \sigmai$ and polynomials 
$\whp_1$ and $\whp_2$ such that $(\forall 
n \geq 0)[\whp_1(n+1) > \whp_1(n) > 0$ and $\whp_2(n+1) > \whp_2(n) > 0]$
and if 
$(x,\omega_s,\omega_{s-1},\dots,\omega_j)$ and 
$(x,\omega'_s,\omega'_{s-1},\dots,\omega'_{j'})$ are 
maximal hard sequences for length $\whp_1(n)$ and $\whp_2(n)$, respectively, 
then for all $y$ and $n$ satisfying $|y|\leq n$ it holds that:
$$y \in \lsigiptwo \iff 
\pair{x,1^n,\omega_s,\omega_{s-1},\dots,\omega_j,\#,\omega'_s,\omega'_{s-1},\dots,
\omega'_{j'},y} \in C.$$}

{\em Proof of Claim 3:} 
Let $B\in \sigmai$ and $\whp$ be a polynomial as defined in Claim 2.  
Let $y$ be a string such that $|y|\leq n$.
According to the definition of $\lsigiptwo$,
$$y \in \lsigiptwo \iff 
(\exists z~ \in \Sigma^{|y|})[\pair{y,z}\notin \lsigipone].$$
Define $\whp_1$ to be a polynomial such that 
$\whp_1(n+1) >\whp_1(n)>0$ and $\whp_1(n)\geq \whp(t(n))$ for all $n$.
In light of Claim 2 we obtain that if 
$(x,\omega_s,\omega_{s-1},\dots,\omega_j)$ is a maximal hard sequence 
for length $\whp_1(n)$ then 
$$y \in \lsigiptwo \iff 
(\exists z \in \Sigma^{|y|})
[\pair{x,1^{\whp_1(n)},\omega_s,\omega_{s-1},\dots,\omega_j,\pair{y,z}}
\not\in B].$$
Set
$B'=\{\pair{x,1^n,\omega_s,\omega_{s-1},\dots,\omega_j,y}\condition
(\exists z \in \Sigma^{|y|})
[\langle x,1^{\whp_1(n)},\omega_s,\omega_{s-1},\dots,\omega_j,
\allowbreak \pair{y,z}
\rangle
\not\in B]\}$.
Clearly $B' \in \sigmaipone$. 
Since $\lsigipone$ is many-one complete for $\sigmaipone$, there exists 
a many-one reduction $g$ from $B'$ to $\lsigipone$, in particular, for all 
$v \in \sigmastar$,
$$v \in B' \iff g(v) \in \lsigipone.$$
Let $q$ be a polynomial such that $|g(v)|\leq q(|v|)$ for all $v$.
Hence 
if the sequence 
$(x,\omega_s,\omega_{s-1},\dots,\omega_j)$ is a maximal hard sequence 
for length $\whp_1(n)$ then 
$$y \in \lsigiptwo \iff 
g(\pair{x,1^n,\omega_s,\omega_{s-1},\dots,\omega_j,y}) \in \lsigipone.$$
Define $\whp_2(n)$ to be a polynomial such that 
$\whp_2(n+1) >\whp_2(n)>0$ and $\whp_2(n)\geq \whp(q(t(\whp_1(n))))$ for all 
$n$.
Applying Claim 2 for the second time we have that 
if 
$(x,\omega'_s,\omega'_{s-1},\dots,\omega'_{j'})$ is a maximal hard 
sequence for length $\whp_2(n)$ then 
$$y \in \lsigiptwo \iff
\pair{x,1^{\whp_2(n)},\omega'_s,\omega'_{s-1},\dots,\omega'_{j'},
g(\pair{x,1^n,\omega_s,\omega_{s-1},\dots,\omega_j,y})} \in B.$$
Define the set  $C$ by   
$\pair{x,1^n,\omega'_s,\omega'_{s-1},\dots,\omega'_{j'},\#,
\omega_s,\omega_{s-1},\dots,\omega_j,y} \in C$ if and only if 
$\pair{x,1^{\whp_2(n)},\omega'_s,\omega'_{s-1},\dots,\omega'_{j'},
g(\pair{x,1^n,\omega_s,\omega_{s-1},\dots,\omega_j,y})} \in B$. 
Note that $C \in \sigmai$.

{\bf Claim 4:} {\em There exists a set 
$D \in \p^{\sigmaipone}$ such that 
for all $y$ and $n$ satisfying $|y|\leq n$ it holds that:
$$y \in \lsigiptwo \iff \pair{x,1^n,y} \in D.
$$}

{\em Proof of Claim 4:}
Let $C\in \sigmai$ and
$\whp_1$ and $\whp_2$ be as in Claim 3. Note that 
$\{\pair{x,1^n,\omega_s,\omega_{s-1},\dots,\omega_j} \condition  
(x,\omega_s,\omega_{s-1},
\dots,\omega_j)$ is a hard string for length
$\whp_1(n) \}$ and 
$\{\pair{x,1^n,\omega_s,\omega_{s-1},\dots,\omega_j} \condition  
(x,\omega_s,\omega_{s-1},
\dots,\omega_j)$ is a hard string for length
$\whp_2(n)\}$
are $\Pi^p_i$ sets.  So, the set of strings 
$\pair{x,1^n,k,\ell}$ such that there exists a hard sequence for
length
$\whp_1(n)$ of length $k$ and there exists a hard sequence for length
$\whp_2(n)$ of length $\ell$ is a $\Sigma_{i+1}^p$ set.

The following $\p^{\sigmaipone}$ algorithm accepts $\pair{x,1^n,y}$ if and only
if
$y
\in \lsigiptwo$: Compute the largest $k$ and $\ell$ such that there exists a 
hard sequence for length
$\whp_1(n)$ of length $k$ and there exists a hard sequence for length
$\whp_2(n)$ of length $\ell$. Then guess a hard sequence
$(x, \omega_s,\omega_{s-1},\dots,\omega_{s-k+2})$ for length $\whp_1(n)$ and
$(x, \omega'_s,\omega'_{s-1},\dots,\omega'_{s-\ell+2})$ for length $\whp_2(n)$ 
and accept if and only if
\\
$\pair{x,1^n,\omega_s,\omega_{s-1},\dots,\omega_{s-k+2},\#,
\omega'_s,\omega'_{s-1},\dots,\omega'_{s-\ell+2},y} \in C$.~\qed


\end{document}